\documentclass[twoside]{article}
\usepackage{qic,epsfig}

\renewcommand{\thefootnote}{\fnsymbol{footnote}}

\begin{document}
\setlength{\textheight}{8.0truein}

\runninghead{A new algorithm for fixed point quantum search}
            {T. Tulsi, L.K. Grover and A. Patel}
\normalsize\textlineskip
\thispagestyle{empty}
\setcounter{page}{1}

\copyrightheading{0}{0}{2005}{000-000}
\vspace*{0.88truein}
\alphfootnote
\fpage{1}

\centerline{\bf A NEW ALGORITHM FOR FIXED POINT QUANTUM SEARCH}
\vspace*{0.37truein}

\centerline{\footnotesize
            TATHAGAT TULSI\footnote{E-mail: tathagat@physics.iisc.ernet.in}}
\vspace*{0.015truein}
\centerline{\footnotesize\it Department of Physics, Indian Institute of Science}
\baselineskip=10pt
\centerline{\footnotesize\it Bangalore-560012, India}
\vspace*{10pt}
\centerline{\footnotesize
            LOV K. GROVER\footnote{E-mail: lkgrover@bell-labs.com}}
\vspace*{0.015truein}
\centerline{\footnotesize\it Bell Laboratories, Lucent Technologies}
\baselineskip=10pt
\centerline{\footnotesize\it Murray Hill, NJ 07974, USA}
\vspace*{10pt}
\centerline{\footnotesize
            APOORVA PATEL\footnote{E-mail: adpatel@cts.iisc.ernet.in}}
\vspace*{0.015truein}
\centerline{\footnotesize\it Centre for High Energy Physics, Indian Institute of Science}
\baselineskip=10pt
\centerline{\footnotesize\it Bangalore-560012, India}
\vspace*{0.225truein}
\publisher{(received date)}{(revised date)}

\vspace*{0.21truein}

\abstracts{
The standard quantum search lacks a feature, enjoyed by many classical
algorithms, of having a fixed point, i.e. monotonic convergence towards the
solution. Recently a fixed point quantum search algorithm has been discovered,
referred to as the Phase-$\pi/3$ search algorithm, which gets around this
limitation. While searching a database for a target state, this algorithm
reduces the error probability from $\epsilon$ to $\epsilon^{2q+1}$ using $q$
oracle queries, which has since been proved to be asymptotically optimal.
A different algorithm is presented here, which has the same worst-case
behavior as the Phase-$\pi/3$ search algorithm but much better average-case
behavior. Furthermore the new algorithm gives $\epsilon^{2q+1}$ convergence
for all integral $q$, whereas the Phase-$\pi/3$ search algorithm requires
$q$ to be $(3^{n}-1)/2$ with $n$ a positive integer. In the new algorithm,
the operations are controlled by two ancilla qubits, and fixed point behavior
is achieved by irreversible measurement operations applied to these ancillas.
It is an example of how measurement can allow us to bypass some restrictions
imposed by unitarity on quantum computing.
}{}{}

\vspace*{10pt}
\vspace*{3pt}
\communicate{}

\vspace*{1pt}\textlineskip
\setcounter{footnote}{0}
\renewcommand{\thefootnote}{\alph{footnote}}

\section{Introduction}

Quantum computing gives us a powerful computational framework, by exploiting
the superposition and entanglement phenomena exhibited by quantum systems.
A famous example of this power is the quantum search algorithm~\cite{grover},
which provides a quadratic speedup over classical search algorithms. This
quantum search algorithm consists of an iterative sequence of selective
inversion and diffusion type operations. Each iteration results in a fixed
rotation (which is a function of the initial error probability) of the
quantum state in a two-dimensional Hilbert space formed by the source and
the target states. If we choose the right number of iteration steps, we stop
very close to the target state, else we keep on going round and round in the
two-dimensional Hilbert space. To perform optimally, therefore, we need to
know the right number of iteration steps, which depends upon the initial
error probability or equivalently the fraction of target states in the
database. We can estimate this fraction to the desired accuracy using various
``amplitude estimation" algorithms~\cite{qaa,qaa1},
but that requires additional queries. When the total number of queries is
large, the additional queries do not matter much, but when the total number
of queries is small, the overhead can be unacceptably large.

In this paper, we address the problem of finding an optimal quantum search
algorithm in situations where, (i) we do not know the initial error
probability (perhaps only its distribution or a bound is known), and
(ii) the expected number of queries is small (so that every additional query
is a substantial overhead). Such situations occur in pattern recognition and
image analysis problems (where each query is a significant cost), and in
problems of error correction and associative memory recall (where the initial
error probability is small but unknown). In these cases, the quantum search
algorithm and its ``amplitude estimation" enhancements are not of much use.
We look for different variations of the quantum search algorithm that ensure
amplitude enhancement, and also outperform classical search algorithms.

The strategy, familiar from classical computation, is to construct an
algorithm that ``converges" towards the target state. That is impossible
to do by iterating a non-trivial unitary transformation; eigenvalues are
of magnitude 1, so the best that can be achieved under such conditions is
a ``limit cycle" and not a ``fixed point". As described above, this is
indeed what happens in case of the quantum search algorithm. To obtain an
algorithm that converges towards a fixed point, some new ingredient is
required, and several possibilities come to mind:
\newline (a) Some property of the current computational state offers an
estimate of the distance to the target state. This estimate can be used as a
parameter to control the extent of the next iterative transformation, e.g. the
Newton-Raphson method to find the zeroes of a function. If such an estimate is
not available, or is too expensive, we have to look for a different method.
\newline(b) Suitably designed but distinct operations are performed at
successive iterations. Such a method can converge towards a fixed point even
using unitary transformations, as exemplified by the recently proposed
Phase-$\pi/3$ search algorithm~\cite{pi3}.
We will describe this in more detail below.
\newline(c) Irreversible damping is introduced in the algorithm without
explicit use of any property of the target state. With the right type of
irreversibility (i.e. when all eigenvalues of the fixed iterative
transformation are less than 1 in magnitude), the algorithm converges,
e.g. the Gauss-Seidel method for solving a set of linear algebraic equations.
Within the framework of quantum computation, such an irreversibility can be
introduced by projective measurement operations, and the algorithm of this
paper falls in this category.

By construction, fixed point algorithms provide two attractive features,
which are not commonplace in generic quantum algorithms:
\newline (1) The initial state is guaranteed to evolve towards the target
state, even when the algorithm is not run to its full completion.
\newline (2) Any errors due to imperfect transformations in earlier
iterations are wiped out by the subsequent iterations, as long as the
state remains in the problem defining space.
\newline These are motivation enough to explore fixed point quantum
algorithms, even if they have other limitations.

Let us consider an unsorted database in which a fraction $f$ of items are
marked, but we don't have precise knowledge of $f$. We run a particular
algorithm which has to return a single item from the database. If the returned
item is a marked one, the algorithm has succeeded, otherwise it is in error.
Without applying any algorithm, if we pick an item at random, then the
probability of error is $\epsilon=1-f$. The goal of the algorithm is to
minimize the error probability, using the smallest number of oracle queries.
If $f$ is sufficiently small, then we can use the optimal quantum search
algorithm to obtain a marked item using $O(1/\sqrt{f})$ queries. A few more
queries to estimate $f$ or to fine-tune the algorithm is not a problem,
because overall we gain a quadratic speed-up compared to the classical case
requiring exhaustive search. But when $f$ is large, the number of oracle
queries is small, and the quantum search algorithm doesn't provide much
advantage---in particular the rotation can overshoot the target state.
In such a situation, a simple classical algorithm (select a random item and
use a query to check if it is a marked one) would outperform the quantum
search algorithm. The same considerations apply to the more general amplitude
amplification algorithms~\cite{qaa,qaa1}.
There the initial quantum state is a unitary operator $U$ applied to a given
source state $|{s}\rangle$, and the probability of getting a target state
after measuring this initial state, $|U_{ts}|^{2}$, is analogous to $f$.
The probability of error, which has to be minimized using the smallest
number of queries, is the probability of getting a non-target state after
measurement, $\epsilon=1-|U_{ts}|^{2}$.

One of us has recently proposed a fixed point quantum search algorithm~\cite{pi3},
which we refer to as the ``Phase-$\pi/3$ search". It shows that by replacing
the selective phase inversions in quantum search algorithm by selective
$\pi/3$-phase shifts, the quantum state monotonically moves closer to the
target state. Remarkably, this convergence is achieved using reversible
unitary transformations and without ever estimating the distance of the
current state from the target state. Explicitly, if the initial error
probability after applying the operator $U$ to the source state $|{s}\rangle$
is $\epsilon$, then the error probability after applying the operator
$UR_{s}^{\pi/3}U^{\dagger}R_{t}^{\pi/3}U$ to $|{s}\rangle$ becomes
$\epsilon^{3}$ (here $R_{s}^{\pi/3}$ and $R_{t}^{\pi/3}$ are $\pi/3$-phase
shift operators for the source and the target state respectively). Recursive
application of this transformation $n$ times, which requires $q_{i}%
=3q_{i-1}+1$ ($q_{1}=1$) oracle queries at the $i^{\mathrm{th}}$ level, makes
the error probability $\epsilon^{3^{n}}$. Thus $\epsilon=0$ is the fixed point
of the algorithm, and the error probability decreases as $\epsilon^{2q+1}$ as
a function of the number of queries $q$. This $\epsilon^{2q+1}$ performance
has been shown to be asymptotically optimal\fnm{a}\fnt{a}{%
Application of a general operator $UR_{s}^{\theta}U^{\dagger}R_{t}^{\phi}U$
to $|{s}\rangle$ changes its component along $|t_\perp\rangle$ by the scale
factor $\big|\exp({i\over2}(\theta-\phi)) - 4\sin{\theta\over2}\sin{\phi\over2}
|U_{ts}|^2\big|$. In the asymptotic limit, $|U_{ts}|\rightarrow1$, and the
scale factor is minimized by the choice $\theta=\phi={\pi\over3}$.}~\cite{jai}.
Note that the best classical algorithm can only decrease the error probability
as $\epsilon^{q+1}$ (not $\epsilon^{q}$, since the last iteration does not
require a query), and thus the Phase-$\pi/3$ search improves the convergence
rate by a factor of 2. Alternative approaches to the Phase-$\pi/3$ search
have been formulated~\cite{jai,tat}, which achieve the same optimal behavior
using different recursive schemes. An important limitation of the Phase-$\pi/3$
search, as well as its alternatives, is that the $\epsilon^{2q+1}$ performance
is obtained only for a restricted number of oracle queries, $q=(3^{n}-1)/2$
(for positive integer $n$).

Here we present a new implementation of fixed point quantum search, that gives
us the optimal $\epsilon^{2q+1}$ performance for all positive integer values of
$q$. It uses a new kind of quantum search, where the oracle and diffusion
operations are controlled in a special way by two ancilla qubits and their
measurement. The same transformation is repeated at every iteration, but since
the transformation is made non-unitary by measurement, the quantum state is
able to monotonically converge towards the target state. Thus the algorithm is
a novel example of how measurement can allow us to bypass restrictions imposed
by unitarity in quantum computing.

In the next section, we begin with a simple scheme for fixed point quantum
search, and then modify it to present the actual algorithm. We analyze the
algorithm in section 3, and discuss its features in section 4.

\section{Algorithm}

To obtain a fixed point quantum search, we have to find an algorithm that
successively decreases the probability of finding a non-target state. Consider
a quantum register whose states encode the items in the database. Let us say
that the initial state of the register is $U\vert{s}\rangle= \sin\theta
\vert{t}\rangle+ \cos\theta\vert{t_{\perp}}\rangle$---an unspecified
superposition of target $\vert{t}\rangle$ and non-target $\vert{t_{\perp}%
}\rangle$ states. (Here, without loss of generality, we have absorbed
arbitrary phases in the definition of $|t\rangle$ and $|t_{\perp}\rangle$.) To
this register, we attach an ancilla bit in the initial state $\vert{0}\rangle$.
Then we perform an oracle query, and flip the ancilla when the register is in
a target state. Now if we measure the ancilla, outcome $1$ tells us that we
are done, and measurement of the register will give us a target state. Outcome
$0$ tells us that the register is in the superposition $|t_{\perp}\rangle$ of
the non-target states. The probability of outcome $0$ is equal to the initial
error probability $\epsilon= \cos^{2}\theta$. To decrease this probability,
we apply the ``diffusion operator" $UI_{s}U^{\dagger}$ to the register,
conditioned on the measurement outcome being $0$. That reflects
$\vert{t_{\perp}}\rangle\vert{0}\rangle$ about $U\vert{s}\rangle\vert{0}
\rangle$ to give the state $\sin2\theta\vert{t}\rangle\vert{0}\rangle+
\cos2\theta\vert{t_{\perp}}\rangle\vert{0}\rangle$. The error probability
has thus decreased by the factor $\cos^{2}2\theta$. Iterating the sequence
of oracle query and diffusion operations $n$ times, the error probability is
reduced to $\cos^{2}\theta \cos^{2n}2\theta$. For $n=1$, the error probability
is $\epsilon(2\epsilon-1)^{2} = 4\epsilon^{3}-4\epsilon ^{2}+\epsilon$.
A comparison of this expression with the corresponding result $\epsilon^{3}$
of the Phase-$\pi/3$ search is shown in Fig.1. We see that for $\epsilon>1/3$,
this simple scheme is better than the Phase-$\pi/3$ search, but it becomes
worse if $\epsilon<1/3$.

\begin{figure}[ptb]
\centerline{\epsfxsize=10cm\epsfbox{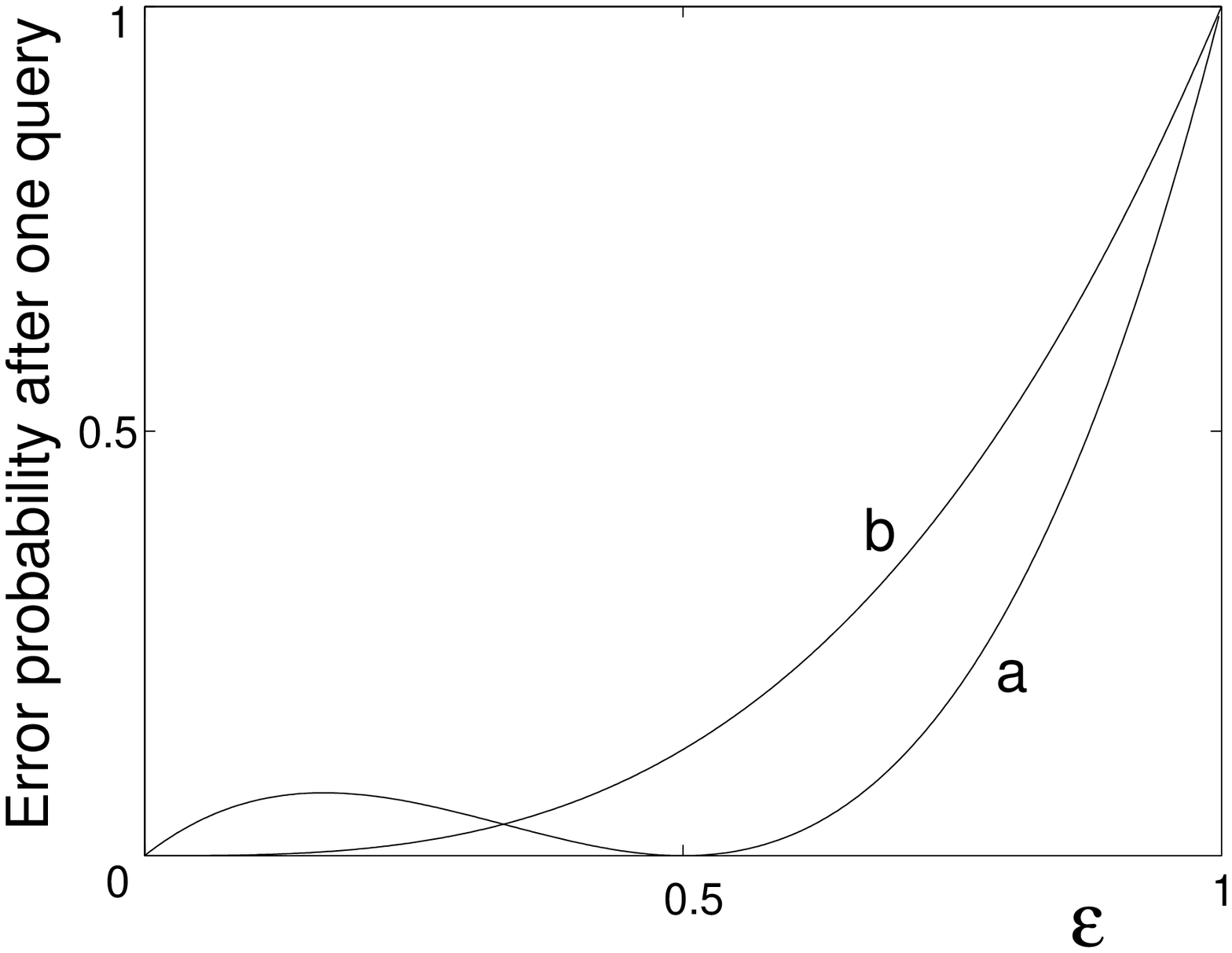}} Figure 1: A comparison of
error probabilities after one oracle query for fixed point quantum search
algorithms. (a) is the result for the simple scheme described at the beginning
of section 2, and (b) is the result for the Phase-$\pi/3$ search algorithm.
\end{figure}

Our goal is to find an algorithm which gives optimal convergence for all
values of $\epsilon$, without knowing any bounds that $\epsilon$ may obey.
Figure 1 shows that, with one iteration of the simple scheme described above,
the error probability monotonically drops from $1$ to $0$ as $\epsilon$
decreases from $1$ to $1/2$. That gives an intuitive idea for getting a better
directed quantum search---somehow set a lower bound of $1/2$ for $\epsilon$,
or equivalently an upper bound of $1/2$ for the fraction of target states $f$.
The natural upper bound for $f$ is $1$, but we can easily make it $1/2$ by
using an extra ancilla in the state $\vert{+}\rangle\equiv(\vert{0}\rangle+
\vert{1}\rangle)/\sqrt{2}$ and performing an oracle query controlled by this
ancilla. This logic suggests the following $q$-iteration algorithm:

\begin{romanlist}
\item Attach two ancilla qubits in the $|0\rangle$ state to the source state
register $|s\rangle$, i.e. $|s\rangle\rightarrow|0\rangle|s\rangle|0\rangle$.
(In what follows, we refer to the former ancilla as ancilla-1 and the latter
ancilla as ancilla-2.)

\item Apply $H\otimes U\otimes I$ to this extended register to prepare the
initial state of the algorithm, $|+\rangle(U|s\rangle) |0\rangle$.

\item Iterate the following two steps, $q$ times:\newline
Step 1: If ancilla-1 is in the state $|1\rangle$, then perform an oracle
query that flips ancilla-2 when the register is in target state.\newline
Step 2: Measure ancilla-2. If the outcome is $1$, the register is certainly
in the target state, so exit the iteration loop. If the outcome is $0$, then
apply the joint diffusion operator $(H\otimes U)I_{0s}(H\otimes U)^{\dagger}$
to the joint state of ancilla-1 and the register.

\item After exiting or completing the iteration loop, stop the quantum
algorithm and measure the register.
\end{romanlist}

The quantum circuit for this algorithm is shown in Fig.2.

\begin{figure}[ptb]
\begin{picture}(300,160)(-50,0)
\put(23,24){\makebox(0,0)[r]{ancilla-2 $\vert{0}\rangle$}}
\put(185,142){\makebox(0,0)[b]{Iterate $q$ times}}
\put(135,147){\vector(-1,0){55}}
\put(235,147){\vector(1,0){55}}
\put(25,24){\line(1,0){45}}
\put(25,75){\line(1,0){15}}
\put(25,125){\line(1,0){15}}
\put(40,65){\framebox(20,20){U}}
\put(40,115){\framebox(20,20){H}}
\put(60,75){\line(1,0){12}}
\put(60,125){\line(1,0){12}}
\put(85,24){\line(1,0){20}}
\put(85,75){\line(1,0){20}}
\put(105,15){\framebox(40,70){Oracle}}
\put(145,24){\line(1,0){20}}
\put(195,22.5){\line(1,0){41}}
\put(195,25.5){\line(1,0){41}}
\put(240,24){\circle{8}}
\put(244,22.5){\line(1,0){41}}
\put(244,25.5){\line(1,0){41}}
\put(250,30){$0$: cont.}
\put(238.5,28){\line(0,1){37}}
\put(241.5,28){\line(0,1){37}}
\put(215,65){\framebox(50,70){Diffusion}}
\put(145,75){\line(1,0){70}}
\put(265,75){\line(1,0){20}}
\put(295,20){\line(0,1){120}}
\put(75,20){\line(0,1){120}}
\put(290,20){\line(1,0){5}}
\put(290,140){\line(1,0){5}}
\put(75,20){\line(1,0){5}}
\put(75,140){\line(1,0){5}}
\put(23,125){\makebox(0,0)[r]{ ancilla-1 $\vert{0}\rangle$}}
\put(23,75){\makebox(0,0)[r]{ register $\vert{s}\rangle$}}
\put(85,125){\line(1,0){36}}
\put(125,125){\circle*{8}}
\put(129,125){\line(1,0){86}}
\put(125,121){\line(0,-1){36}}
\put(265,125){\line(1,0){20}}
\put(165,10){\framebox(30,20)}
\qbezier(169,16)(180,30)(191,16)
\put(180,11){\vector(1,4){4.5}}
\put(250,3){$1$: exit}
\put(195,17){\line(1,0){126}}
\put(195,14){\line(1,0){126}}
\put(325,15.5){\circle*{8}}
\put(323.5,19){\line(0,1){46}}
\put(326.5,19){\line(0,1){46}}
\put(298,75){\line(1,0){12}}
\put(310,66){\framebox(30,20)}
\qbezier(314,71)(325,85)(336,71)
\put(325,66){\vector(1,4){4.5}}
\end{picture}
\par
Figure 2: Quantum circuit for our $q$-iteration algorithm.
\end{figure}

\section{Analysis}

Let us analyze the algorithm step by step. The initial state is
\begin{equation}
\vert{\psi_{i}}\rangle= (H\otimes U\otimes I)\vert{0}\rangle\vert{s}%
\rangle\vert{0}\rangle= \left( \frac{\vert{0}\rangle+ \vert{1}\rangle}%
{\sqrt{2}}\right) \Big(\sin\theta\vert{t}\rangle+\cos\theta\vert{t_{\perp}%
}\rangle\Big)\vert{0}\rangle~.
\end{equation}
The initial error probability is $\epsilon=\cos^{2}\theta$, and the initial
success probability is $f=1-\epsilon=\sin^{2}\theta$. We work in the joint
search space of ancilla-1 and the register, denoted by the subscript $j$,
where all the states $\vert{0}\rangle\vert{t}\rangle, \vert{0}\rangle
\vert{t_{\perp}}\rangle, \vert{1}\rangle\vert{t_{\perp}}\rangle$ act as
non-target states. Only the state $\vert{t}_{j}\rangle\equiv\vert{1}%
\rangle\vert{t}\rangle$ acts as the target state in this joint search space,
and let us represent the superposition of all non-target states by
$\vert{t^{\prime}_{j}}\rangle$. In the joint search space, the initial state
is
\begin{equation}
\vert{\psi_{i}}\rangle= \left( \frac{\sin\theta}{\sqrt{2}}\right) \vert{t_{j}%
}\rangle\vert{0}\rangle+ \frac{1}{N}\vert{t^{\prime}_{j}}\rangle\vert
{0}\rangle~,
\end{equation}
where the unit vector $|t^{\prime}_{j}\rangle$ is
\begin{equation}
\vert{t^{\prime}_{j}}\rangle= N\left( \frac{\sin\theta}{\sqrt{2}}\vert
{0}\rangle\vert{t}\rangle+\frac{\cos\theta}{\sqrt{2}}\vert{0}\rangle
\vert{t_{\perp}}\rangle+\frac{\cos\theta}{\sqrt{2}}\vert{1}\rangle
\vert{t_{\perp}}\rangle\right)  ~,
\end{equation}
and the normalization factor $N$ is $\left( \cos^{2}\theta+ \frac{1}{2}%
\sin^{2}\theta\right) ^{-1/2} = \sqrt{2/(1+\epsilon)}$. For later reference,
note that the error probability after measuring the joint state
$\vert{t^{\prime}_{j}}\rangle$ (i.e. the probability of finding the register
in the non-target state $\vert{t_{\perp}}\rangle$) is
\begin{equation}
\vert\langle0 \vert\langle t_{\perp}\vert t^{\prime}_{j}\rangle\vert^{2} +
\vert\langle1 \vert\langle t_{\perp}\vert t^{\prime}_{j}\rangle\vert^{2} =
N^{2}\cos^{2}\theta= N^{2}\epsilon~.
\end{equation}

Step 1 of the algorithm, using an oracle query, flips ancilla-2 when the joint
register state is $\vert{t_{j}}\rangle$. In step 2, we measure ancilla-2. If
the outcome is $1$ then we stop the algorithm, because the register is in the
target state. The probability of getting $1$ is $\sin^{2}\theta/2=f/2$---we
have effectively put an upper bound of $1/2$ on the success probability using
ancilla-1. If the outcome is $0$, which has probability $1/N^{2}$, then the
joint state is $\vert{t^{\prime}_{j}}\rangle$. In this case, we apply the
joint diffusion operation using the joint source state $\vert{s_{j}}%
\rangle\equiv\vert{0}\rangle\vert{s}\rangle$. The joint diffusion operation is
a reflection about $(H\otimes U)|0\rangle|s\rangle\equiv U_{j}|s_{j}\rangle$,
in the two-dimensional Hilbert space orthogonally spanned by $\vert{t_{j}%
}\rangle$ and $\vert{t^{\prime}_{j}}\rangle$ as shown in Fig.3. The state
$U_{j}\vert{s_{j}}\rangle$ makes an angle $\theta_{j}$, defined by $\sin
^{2}\theta_{j} = \sin^{2}\theta/2$, with the state $\vert{t^{\prime}_{j}%
}\rangle$. So reflecting $\vert{t^{\prime}_{j}}\rangle$ about $U_{j}%
\vert{s_{j}}\rangle$ gives us a state that makes an angle $2\theta_{j}$ with
$\vert{t^{\prime}_{j}}\rangle$. Its component in $\vert{t^{\prime}_{j}}%
\rangle$-direction is $\cos2\theta_{j} = 1-2\sin^{2}\theta_{j} = 1-\sin
^{2}\theta= \epsilon$. Thus the joint diffusion operation produces the final
state
\begin{equation}
\vert{\psi_{f}}\rangle= U_{j} I_{s_{j}} U_{j}^{\dagger}|t^{\prime}_{j}\rangle=
\sqrt{1-\epsilon^{2}}\vert{t_{j}}\rangle+\epsilon\vert{t^{\prime}_{j}}%
\rangle~.
\end{equation}
After measuring $\vert{\psi_{f}}\rangle$, the probability of getting
$\vert{t^{\prime}_{j}}\rangle$ is $\epsilon^{2}$, so the total probability of
getting $\vert{t^{\prime}_{j}}\rangle$ after one iteration is $N^{-2}%
\epsilon^{2}$. Iterating the algorithm will keep on decreasing this
probability by a factor of $\epsilon^{2}$ at each iteration, and after $q$
iterations it will become $N^{-2}\epsilon^{2q}$. So the net error probability
after $q$ iterations is (cf. Eq.(4))
\begin{equation}
\epsilon_{q} = N^{2}\epsilon(N^{-2}\epsilon^{2q}) = \epsilon^{2q+1} ~,
\end{equation}
which agrees exactly with the corresponding result for the Phase-$\pi/3$
search.

\begin{figure}[ptb]
\begin{picture}(500,360)(0,-40)
\renewcommand{\qbeziermax}{60}
\put(15,15){\vector(1,0){65}}
\put(15,15){\vector(0,1){65}}
\put(160,265){\vector(1,0){65}}
\put(15,265){\vector(1,0){65}}
\put(305,265){\vector(1,0){65}}
\put(305,15){\vector(1,0){65}}
\put(160,140){\vector(1,0){65}}
\put(160,15){\vector(1,0){65}}
\put(160,265){\vector(0,1){65}}
\put(15,265){\vector(0,1){65}}
\put(305,265){\vector(0,1){65}}
\put(305,15){\vector(0,1){65}}
\put(160,140){\vector(0,1){65}}
\put(160,15){\vector(0,1){65}}
\qbezier(15,265)(30,290.6)(47.5,321.3)
\put(47.5,321.3){\vector(0,1){2}}
\qbezier(160,265)(183.7,283.4)(211.4,304.8)
\put(211.4,304.8){\vector(1,1){2}}
\qbezier(305,265)(328.7,283.4)(356.4,304.8)
\put(356.4,304.8){\vector(1,1){2}}
\put(13,335){\makebox(0,0)[r]{$\vert{t}\rangle$}}
\put(95,263){\makebox(0,0)[tr]{$\vert{t_\perp}\rangle$}}
\put(51,331){\makebox(0,0)[tl]{$U\vert{s}\rangle$}}
\put(240,263){\makebox(0,0)[tr]{$\vert{t'_{j}}\rangle$}}
\put(158,335){\makebox(0,0)[r]{$\vert{t_{j}}\rangle$}}
\put(216,314.5){\makebox(0,0)[tl]{$U_{j}\vert{s_{j}}\rangle$}}
\put(400,263){\makebox(0,0)[tr]{$\vert{t'_{j}}\rangle\vert{0}\rangle$}}
\put(303,335){\makebox(0,0)[r]{$\vert{t_{j}}\rangle\vert{1}\rangle$}}
\put(361,314.5){\makebox(0,0)[tl]{$U_{j}\vert{s_{j}}\rangle$}}
\qbezier(25,282.3)(36.65,277.5)(35,265)
\put(36.65,277.5){$\theta$}
\qbezier(175.8,277.25)(178.92,271.47)(180,265)
\put(182,271.47){$\theta_{j}$}
\qbezier(320.8,27.25)(323.92,21.47)(325,15)
\put(327,21.47){$\theta_{j}$}
\qbezier(320.8,277.25)(323.92,271.47)(325,265)
\put(327,271.47){$\theta_{j}$}
\qbezier[25](305,15)(328.7,33.4)(356.4,54.8)
\put(356.4,54.8){\vector(1,1){2}}
\qbezier[25](160,15)(183.7,33.4)(211.4,54.8)
\put(211.4,54.8){\vector(1,1){2}}
\qbezier(160,15)(167.5,44.04)(176.25,77.93)
\put(176.25,77.93){\vector(0,1){2}}
\qbezier(160,140)(167.5,169.04)(176.25,202.93)
\put(176.25,202.93){\vector(0,1){2}}
\qbezier[25](160,140)(183.7,158.4)(211.4,179.8)
\put(211.4,179.8){\vector(1,1){2}}
\put(305,17){\vector(1,0){65}}
\put(295,145){\framebox(85,55)}
\put(297,198){\makebox(0,0)[tl]{Measure ancilla}}
\put(297,182){\makebox(0,0)[tl]{If outcome is}}
\put(297,170){\makebox(0,0)[tl]{ $1$: exit loop}}
\put(297,158){\makebox(0,0)[tl]{ $0$: continue}}
\put(370,13){\makebox(0,0)[tl]{$\vert{t'_{j}}\rangle$}}
\put(303,80){\makebox(0,0)[br]{$\vert{t_{j}}\rangle$}}
\put(361,64.5){\makebox(0,0)[tl]{$U_{j}\vert{s_{j}}\rangle$}}
\put(225,13){\makebox(0,0)[tl]{$\vert{t'_{j}}\rangle$}}
\put(158,80){\makebox(0,0)[br]{$\vert{t_{j}}\rangle$}}
\put(216,64.5){\makebox(0,0)[tl]{$U_{j}\vert{s_{j}}\rangle$}}
\put(225,138){\makebox(0,0)[tl]{$\vert{t'_{j}}\rangle\vert{0}\rangle$}}
\put(158,205){\makebox(0,0)[br]{$\vert{t_{j}}\rangle\vert{1}\rangle$}}
\put(216,189.5){\makebox(0,0)[tl]{$U_{j}\vert{s_{j}}\rangle$}}
\put(95,13){\makebox(0,0)[tr]{$\vert{t_\perp}\rangle$}}
\put(13,85){\makebox(0,0)[r]{$\vert{t}\rangle$}}
\put(126,-2){\dashbox(267,223)}
\put(177,203){\makebox(0,0)[bl]{$U_{j}I_{s_{j}}U_{j}^{\dagger}\vert{t'_{j}}\rangle$}}
\put(177,78){\makebox(0,0)[bl]{$U_{j}I_{s_{j}}U_{j}^{\dagger}\vert{t'_{j}}\rangle$}}
\put(105,300){$\Longrightarrow$}
\put(105,288){Initial}
\put(105,276){state}
\put(245,294){\makebox(0,0)[tl]{ Oracle}}
\put(250,297){\vector(1,0){30}}
\put(337,255){\vector(0,-1){45}}
\put(337,130){\vector(0,-1){30}}
\put(285,42){\vector(-1,0){30}}
\put(192,105){\line(0,1){8}}
\put(192,124){\vector(0,1){8}}
\put(250,172){\vector(1,0){30}}
\put(192,120){\makebox(0,0){Oracle}}
\put(246,38){\makebox(0,0)[tl]{Diffusion}}
\qbezier(165,34.36)(175.81,27.25)(180,15)
\put(170,34){$2\theta_{j}$}
\put(170,159){$2\theta_{j}$}
\qbezier(165,159.36)(175.81,152.25)(180,140)
\put(339,115){\makebox(0,0)[l]{$\vert{t'_{j}}\rangle$}}
\qbezier(15,15)(33.97,38.238)(56.11,65.35)
\qbezier(15,15)(19.74,44.62)(25.28,79.18)
\put(56.11,65.35){\vector(1,1){2}}
\put(25.28,79.18){\vector(0,1){2}}
\put(56,62){\makebox(0,0)[tl]{$\vert{t'_{j}}\rangle$}}
\put(29,76){\makebox(0,0)[bl]{$U_{j}I_{s_{j}}U_{j}^{\dagger}\vert{t'_{j}}\rangle$}}
\put(102,48){$\Longleftarrow$}
\put(102,55){exit}
\put(50,10){\vector(0,-1){20}}
\put(10,-20){Measure register}
\put(92,172){\vector(1,0){30}}
\put(90,174){\makebox(0,0)[br]{Iterate the loop}}
\put(82,162){\makebox(0,0)[br]{$q-1$ times}}
\end{picture}
Figure 3: Step-by-step quantum state evolution in our $q$-iteration algorithm.
The double arrows relate the equivalent states in the original search space
(left) and the joint search space (right).
\end{figure}

The above analysis allows us to also deduce the following features:
\newline(1) $\epsilon=1$ can be made a fixed point of the algorithm, instead
of $\epsilon=0$, by effectively interchanging the roles of $|t\rangle$ and
$|t_{\perp}\rangle$. This is achieved by flipping ancilla-2, only when the
joint state is $|1\rangle|t_{\perp}\rangle$. Then the probability of finding
the register in the state $|t\rangle$, after $q$ iterations, becomes
$(1-\epsilon)^{2q+1}$. Note that the same behavior can be obtained in
the Phase-$\pi/3$ search, by replacing either $R_{t}^{\pi/3}$ with
$R_{t}^{-\pi/3}$ or $R_{s}^{\pi/3}$ with $R_{s}^{-\pi/3}$. This $\epsilon=1$
fixed point can be useful in situations where certain target states are to be
avoided, e.g. in collision problems.
\newline(2) When we use fixed point quantum search to locate the target state
in a database, the initial error probability is $\epsilon=1-f$. The number
of oracle queries required to reduce this probability to $o(1)$ obeys
\begin{equation}
(1-f)^{2q+1} = o(1) ~~\mathop{\longrightarrow}\limits^{f~small}~~ e^{-(2q+1)f}
= o(1) ~.
\end{equation}
Thus we need $q=O(1/f)$ oracle queries to find the target state reliably.
The same is true of the Phase-$\pi/3$ search as well as its
alternatives~\cite{jai,tat}. This scaling of fixed point quantum search is
clearly inferior to the $O(1/\sqrt{f})$ scaling of the quantum search
algorithm. Still, as discussed in the introduction, fixed point quantum
search can be useful in situations where $f$ is unknown and large.
\newline(3) It is possible to stick to unitary operations throughout the
algorithm and postpone measurement till the very end. In such a scenario,
the unmeasured ancilla-2 has to control the diffusion operation and all the
subsequent iterations (i.e. they are executed only when ancilla-2 is in the
$|0\rangle$ state), and it cannot be reused in the iteration loop. We need
a separate ancilla-2 for every oracle query, to ensure that once the states
$|t_{j}\rangle$ and $|t^{\prime}_{j}\rangle$ are separated by an oracle query
in an iteration, they are not superposed again by subsequent iterations.
The whole set of $q$ ancilla-2 can be measured after the iteration loop,
in a sequence corresponding to the iteration number, to determine the target
state. In this version, the unitary transformation is different for each
iteration (because each iteration involves a different ancilla-2 and
different controls), and the unitary iterations converge to a fixed point.

\section{Comparison}

Now we can point out some advantageous features of our algorithm compared to
other algorithms~\cite{pi3,jai} for fixed point quantum search. It is also
instructive to compare our results with a simple classical algorithm, where
each iteration consists of picking an item randomly from the database and
testing it by an oracle query, giving an error probability $\epsilon^{q+1}$
after $q$ iterations.

\textbf{Real variables:} No complex numbers appear in our algorithm, unless
they are part of the operator $U$. This increases the ease as well as the
possibilities for physical implementation of the algorithm. In particular,
the algorithm can be realized using classical waves (with individual states
represented as orthogonal wave modes), as is the case for the quantum search
algorithm~\cite{wave}.

\textbf{Allowed values of $q$:} A practical criterion for stopping the
iterative algorithm would be that the error probability becomes smaller than
some predetermined threshold $\epsilon_{\mathrm{th}}$. Provided we have an
upper bound, $\epsilon\le\epsilon_{\mathrm{up}} < 1$, we can guarantee
convergence by choosing $\epsilon_{\mathrm{up}}^{2q+1}\le\epsilon_{\mathrm{th}}$.
In our algorithm $2q+1$ can take all odd positive integer values, while in
the Phase-$\pi/3$ search (or its alternate versions) $2q+1$ can only take a
restricted set of values of the form $3^{n}$ (for positive integer $n$).
Thus we can use $q_{\mathrm{an}}=\lceil{\frac{1 }{2}} \lceil{\frac{\log
\epsilon_{\mathrm{th}} }{\log\epsilon_{\mathrm{up}}}} - 1 \rceil\rceil$,
while the Phase-$\pi/3$ search requires $q_{\pi/3}={\frac{1 }{2}}( 3^{\lceil
\log_{3}(\log\epsilon_{\mathrm{th}}/\log\epsilon_{\mathrm{up}}) \rceil} - 1 )$
($=1,4,13,40,121,364,1093,\ldots)$. Thus we can save a sizable number of
oracle queries---up to a factor of $3$---compared to the Phase-$\pi/3$
search, especially when $\log_{3}(\log\epsilon_{\mathrm{th}}/\log
\epsilon_{\mathrm{up}})$ slightly exceeds an integer. Moreover, the simple
classical algorithm needs $q_{\mathrm{cl}}=\lceil{\frac{\log\epsilon
_{\mathrm{th}} }{\log\epsilon_{\mathrm{up}}}} - 1 \rceil$ oracle queries,
which is always more than that for our algorithm (by about a factor of $2$)
but can be less than that for the Phase-$\pi/3$ search (by up to a factor of
$2/3$).

\textbf{Deterministic vs. probabilistic:} Although our complete algorithm is
probabilistic, whenever we exit the iteration loop after finding ancilla-2
in the state $|1\rangle$, we obtain a deterministic result for the target
state. The total probability of this deterministic result, over $q$ iterations
of our algorithm, is
\begin{eqnarray}
P_{1}(q) &=& \left(\frac{1-\epsilon}{2}\right)
	   + \left(\frac{1+\epsilon}{2}\right) (1-\epsilon^{2})
             \sum_{k=2}^{q}\epsilon^{2k-4} \nonumber\\
	 &=& \left(\frac{1-\epsilon}{2}\right)
           + \left(\frac{1+\epsilon}{2}\right) (1-\epsilon^{2q-2})
         ~=~ 1-\epsilon^{2q-2} \left(\frac{1+\epsilon}{2}\right) ~.
\end{eqnarray}
Only the last measurement of the register gives a probabilistic result
for the target state, with probability ${\frac{1}{2}}\epsilon^{2q-2}
(1+\epsilon-2\epsilon^{3})$. The fact that a major fraction
(asymptotically all) of the total $1-\epsilon^{2q+1}$ success probability
gives a deterministic result can be a useful feature in some applications.
Such a deterministic feature is not present in the Phase-$\pi/3$ search,
but it exists in its alternate version~\cite{jai}.

\textbf{Average number of oracle queries:} The Phase-$\pi/3$ search always
requires the same number of queries as the number of iterations, because it
has to go through all the iterations. On the other hand our algorithm can
terminate at various times depending on the outcome of the ancilla-2
measurements. The average number of queries is always less than the number
of iterations because our algorithm generally exits the iteration loop before
completing it. We calculate the average number of oracle queries for our
algorithm as in Eq.(8),
\begin{eqnarray}
\overline{q}_{\mathrm{an}} &=& \left(\frac{1-\epsilon}{2}\right)
  + \left(\frac{1+\epsilon}{2}\right) (1-\epsilon^{2})\sum_{k=2}^{q}
  k\epsilon^{2k-4} + q \left(\frac{1+\epsilon}{2}\right) \epsilon^{2q-2}
  \nonumber\\
  &=& \left(\frac{1-\epsilon}{2}\right) + \left(\frac{1+\epsilon}{2}\right)
      \left(1+\frac{1-\epsilon^{2q-2}}{1-\epsilon^{2}}\right)
  ~=~ 1+\frac{1-\epsilon^{2q-2}}{2(1-\epsilon)} ~.
\end{eqnarray}
For the simple classical algorithm, the average number of oracle queries in
case of $2q$ iterations is
\begin{equation}
\overline{q}_{\mathrm{cl}} = (1-\epsilon)\sum_{k=1}^{2q}k\epsilon^{k-1}
  + 2q\epsilon^{2q} = \frac{1-\epsilon^{2q}}{1-\epsilon} ~.
\end{equation}
In general, $2\overline{q}_{\mathrm{an}}>\overline{q}_{\mathrm{cl}}$, so our
quantum algorithm can save at most a factor of $2$ in the number of oracle
queries compared to the simple classical algorithm. For $q=1$, i.e. when the
error probability is reduced to $\epsilon^{3}$, $\overline{q}_{\mathrm{an}}=1$
is always better than $\overline{q}_{\mathrm{cl}}=1+\epsilon$. For $q>1$, our
quantum algorithm provides an advantage only for $\epsilon>\epsilon_{a}$,
where $\epsilon_{a}$ satisfies $2\epsilon_{a}+\epsilon_{a}^{2q-2}%
-2\epsilon_{a}^{2q}=1$ and is very close to but always less than $1/2$. For
$\epsilon<\epsilon_{a}$, the simple classical algorithm requires less number
of oracle queries than our quantum algorithm, but in this range $\overline
{q}<2$ and hence the disadvantage is not significant. To illustrate this
behavior, we have plotted the average number of oracle queries as a function
of $\epsilon$ in Fig.4, for the three algorithms when $q=4$.

\begin{figure}[ptb]
\centerline{\epsfxsize=10cm\epsfbox{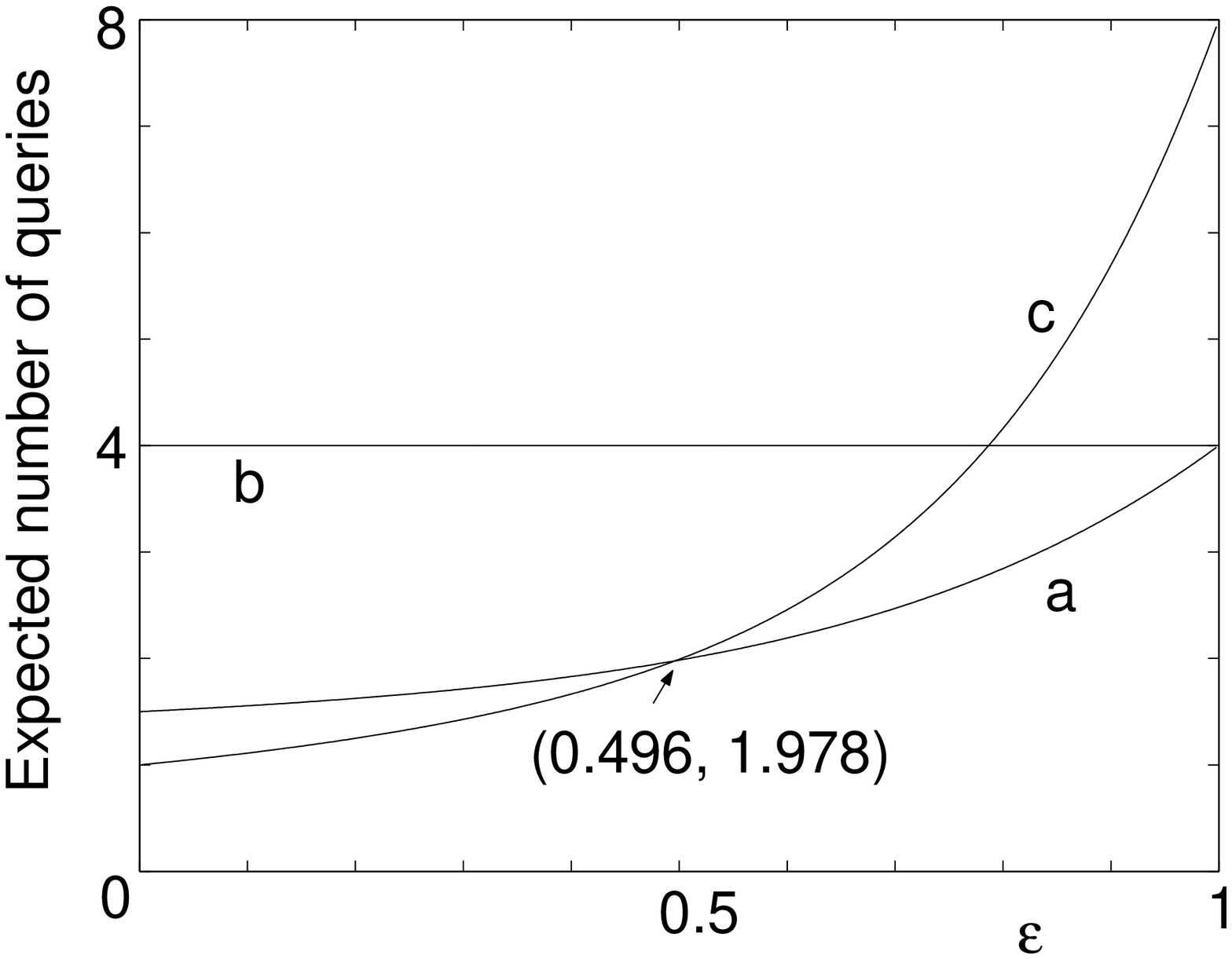}} Figure 4: Comparison of the
average number of oracle queries for fixed point search algorithms, when $q=4$.
(a) is for our algorithm, (b) is for the Phase-$\pi/3$ search algorithm, and
(c) is for the simple classical algorithm.
\end{figure}

\textbf{General ancilla-1 state:} In the algorithm discussed so far, using
ancilla-1 in the initial state $|+\rangle$, we effectively halved the fraction
of target states. More generally, we can make the fraction of target states
smaller by a factor $r$, by choosing ancilla-1 in the initial state
$\sqrt{1-r}|{0}\rangle+\sqrt{r}|{1}\rangle$. The corresponding joint
unitary operator $U_{j}$ becomes $R\otimes U$, where $R$ transforms
$|{0}\rangle$ to $\sqrt{1-r}|{0}\rangle+\sqrt{r}|{1}\rangle$. It can be
easily shown that after $q$ iterations of the algorithm, the probability of
getting a non-target state becomes $\epsilon\big(1-2r(1-\epsilon)\big)^{2q}$.
Also, the average number of oracle queries becomes (cf. Eq.(9))
\begin{eqnarray}
\overline{q}_{\mathrm{an}}(r) &=& r(1-\epsilon) + (1-r(1-\epsilon))
  \left(1+\frac{1-(1-2r(1-\epsilon))^{2q-2}}{1-(1-2r(1-\epsilon))^{2}}\right)
  \nonumber\\
  &=& 1+\frac{1-(1-2r(1-\epsilon))^{2q-2}}{4r(1-\epsilon)} ~.
\end{eqnarray}

\textbf{Optimization of worst-case and average-case behavior:} For a given
initial error probability $\epsilon$, the optimal worst-case algorithm
provides the fastest decrease of the error probability as a function of $q$,
while the optimal average-case algorithm yields the smallest value of 
$\overline{q}$. The above results give us the following solutions:
\newline(a) For $r>1/2$, the error probability vanishes for a non-zero value
of $\epsilon$, $\epsilon_{0}=1-1/(2r)$. So if $\epsilon\in(0,\frac{1}{2}]$
is known, a single oracle ($q=1$) can find a target state with certainty.
\newline(b) For $r>1/2$, there is a particular value of $\epsilon$,
given by $\epsilon_{rl}=\frac{2r-1}{2r+1}\in(0,\frac{1}{3}]$, above which
$|1-2r(1-\epsilon)|<\epsilon$. Thus, if we have a lower bound on $\epsilon$,
choosing $r$ to make it equal to $\epsilon_{rl}$ gives an algorithm that
converges faster than $\epsilon^{2q+1}$, i.e. the Phase-$\pi/3$ search.
It has been shown~\cite{jai,tat} that for one iteration, the same behavior
can be obtained using only one ancilla. In absence of a lower bound on
$\epsilon$, our algorithm with $r=1/2$ remains the best bet.
\newline(c) The optimal average case algorithm is easily obtained by taking
the limit $q\rightarrow\infty$.
\begin{equation}
\overline{q}^{\infty}_{\mathrm{an}}(r) = 1+\frac{1}{4r(1-\epsilon)} ~,~~
\overline{q}^{\infty}_{\mathrm{cl}} = \frac{1}{1-\epsilon} ~.
\end{equation}
Clearly the smallest value for $\overline{q}$ is obtained for $r=1$,
which is the simple scheme without ancilla-1 described at the beginning
of section 2. It can give up to a factor of 4 improvement over the simple
classical algorithm, when $\epsilon>1/4$. The simple classical algorithm
is better for $\epsilon<1/4$, but in this range $\overline{q}<4/3$ and
the difference is rather small.

\textbf{No. of ancilla states:} Our algorithm uses only two ancilla qubits.
The Phase-$\pi/3$ search needs a six-state ancilla to obtain the phase
transformations from the standard binary oracle $O_{f}:|{x}\rangle|{a}%
\rangle\rightarrow|{x}\rangle|{a\oplus_6 f(x)}\rangle$. Its alternative
versions~\cite{jai,tat} use one ancilla qubit per recursion step, and so
require $n$ ancilla qubits where $2q+1=3^{n}$. Our algorithm is therefore
quite economical in the number of ancilla states required.

A related issue is whether or not the algorithm can be improved by POVM
measurements. Since POVM measurements are equivalent to projective
measurements in an enlarged Hilbert space (which can be obtained by
adding extra ancilla qubits), and our algorithm has already achieved
the optimal asymptotic performance with only two ancilla qubits,
there is not much left for POVM measurements to improve up on.

\section{Outlook}

We have presented a new algorithm for fixed point quantum search using
irreversible measurement operations, which is superior to the Phase-$\pi/3$
search in several respects. Fixed point quantum search algorithms provide new
techniques for driving a quantum state towards a specific target. Although
they are no match for the optimal quantum search algorithm for large number
of queries, i.e. $O(1/f)$ compared to $O(1/\sqrt{f})$ for small $f$, they
are still better than the best classical search algorithm by a factor of 2.
The potential applications of these algorithms are in problems with small
but unknown initial error probability and limited number of oracle queries,
where they assure monotonic power-law convergence while the optimal quantum
search algorithm is uncertain in improvement. This has been illustrated
in case of correction of systematic quantum errors~\cite{pi3,reichardt}.

The concept of ancilla-controlled quantum search presented here can be
generalized to make the ancilla transformations iteration dependent,
in a manner reminiscent of local adiabatic quantum algorithms,
and that will be presented in a forthcoming paper~\cite{tat}.

\section*{Acknowledgments}

T. Tulsi acknowledges the fellowship from Council of Scientific and
Industrial Research, India, during the course of this work.

\newpage
\nonumsection{References}

\end{document}